\documentclass[12pt]{article}
\usepackage[centertags]{amsmath}
\usepackage{graphicx}
\usepackage[]{epsfig}
\usepackage{color}
\newcommand{\beq}{\begin{equation}}
\newcommand{\eeq}{\end{equation}}
\newcommand{\ba}{\begin{eqnarray}}
\newcommand{\ea}{\end{eqnarray}}
\newcommand{\be}{\begin{equation}}
\newcommand{\ee}{\end{equation}}

\newcommand{\bea}{\begin{eqnarray}}
\newcommand{\eea}{\end{eqnarray}}

\newcommand{\eme}{\mathpzc{m}}
\usepackage[centertags]{amsmath}
\usepackage{amsfonts}

\normalbaselines
\DeclareFontFamily{OT1}{pzc}{}
\DeclareFontShape{OT1}{pzc}{m}{it}{<-> s * [1.100] pzcmi7t}{}
\DeclareMathAlphabet{\mathpzc}{OT1}{pzc}{m}{it}
\begin{document}
\title{Bosonization of fermions coupled to topologically massive gravity}
\author{Eduardo Fradkin$^a$,  Enrique Moreno$^b$  and
Fidel~A.~Schaposnik$^c$\thanks{Also at CICBA.} \\ \vspace{0.2 cm}\\
{\normalsize \it $^a$Department of Physics and Institute for Condensed Matter Theory,}\\ {\normalsize \it University of Illinois at Urbana-Champaign,}\\{\normalsize \it
 1110 West Green Street, Urbana, Illinois, 61801-3080, USA.} \\
\\
{\normalsize \it $^b$Department of Physics, Northeastern University}\\ {\normalsize \it Boston, MA 02115,
USA.} \\
\\
{\normalsize \it $^c$\it Departamento de F\'\i sica, Universidad Nacional de La Plata}\\ {\normalsize \it Instituto de F\'\i sica La Plata}\\ {\normalsize\it C.C. 67, 1900 La Plata, Argentina.}}

\date{\today}

\maketitle
\begin{abstract}
 We establish a   duality between   massive fermions   coupled to topologically massive gravity (TGM) in $d=3$ space-time dimensions and a purely gravity theory which also will turn out to be a TGM theory but with different parameters: the original graviton mass in the  TGM theory coupled to fermions picks-up a contribution from fermion bosonization. We obtain explicit bosonization rules for the  fermionic currents and for the energy-momentum-tensor showing that the identifications do not depend explicitly on the parameters of the theory.  These results are the gravitational analog of the results for $2+1$ Abelian and non-Abelian bosonization in flat space-time.
\end{abstract}

Bosonization  of fermion models is strongly connected to the existence of quantum anomalies  associated to symmetries that exist at the classical level.  Indeed, the well-honored fermions-boson duality
  in   $d=2$ space-time dimensions   \cite{Coleman} can be reproduced within the path-integral approach once  the chiral anomaly affecting the fermionic measure \cite{F1} is taken into account \cite{Roskies,Naon}. In the case of $d=3$ fermions coupled to Abelian and non-Abelian gauge fields  it is  the   parity anomaly that plays a central  role unveiling the occurrence of a  Chern-Simons term \cite{DJT,Redlich}  in the bosonic
dual \cite{Marino,FS,LGMNS}.

Interestingly enough,  when fermions are coupled to a gravitational field in odd-dimensional spaces the parity anomaly also induces   a Chern-Simons term which can be written in terms of the spin connection or in terms of the Christoffel connection \cite{AG,Ojima,Kulikov}. Recently this fact  has been exploited to simulate the effects of crystal defects on the electronic degrees of freedom of topological insulators in condensed matter physics. This systems exhibit physical phenomena related to  parity breaking that  take place  in terms of  Dirac fermion models in  $2+1$ space-time dimensions in the presence of gravitational sources \cite{Fr1,Le2}.

~

In view of the discussion above,  it is natural to consider the possibility of finding a bosonization recipe connecting fermions with gravitons analogous to those arising  for fermionic models coupled to gauge fields  in flat space. However, contrary to the standard Fermi-Bose duality, in this case the Dirac fermions will be coupled not to background gravitational sources but to a {\em dynamical} gravity theory,  Topologically Massive Gravity (TGM) theory.  Dualities involving    gravity theories and  fermion models  have recently been discussed within   the gauge/gravity correspondence but in that case the gravity theory (a Vasiliev high-spin theory \cite{V}) is defined in an AdS$_4$ bulk while the fermion dual - consisting of $k$ fermions  coupled to a $U(k)_N$ Chern-Simons term - lives in the three-dimensional boundary of AdS$_4$ \cite{A}. In fact, since also a bosonic theory coupled to a $U(N)_k$ Chern-Simons gauge theory defined on the same boundary can be seen to be dual to Vasiliev theory in the AdS$_4$ bulk, it has been conjectured that these bosonic and fermionic theories are dual to each other in the large $N$ limit.

The aim of this work is to establish a   duality between   fermions   coupled to topologically massive gravity (TGM) in $d=3$ space-time dimensions and a purely gravity theory, which also will turn out to be a TGM theory but with
different parameters.

\subsection*{The action}
Our model will be defined on a three-dimensional space-time manifold $M^3$ with local coordinates $x^\mu$ ($\mu=0,1,2$) and a metric of Minkowski signature. Vierbeins (or, rather, dreibeins in $2+1$ dimensions) are denoted as $e^a_\mu$ where $``a"$ is the frame index ($a=0,1,2$).
The  spin  connection  $\omega_\mu$,
\be
\omega_\mu = -\frac14 \omega_{ab\mu} \gamma^a \gamma^b = -\frac{i}4  \omega_{ab\mu}\epsilon^{abc}\gamma^c
\ee
 is an $SO(2,1)$-valued 1-form on the manifold $M^3$ satisfying the standard relation with the dreibeins
\be
\omega_{ab\mu} = e_{a\nu} \, e^\nu_{b\,; \mu}
\ee
where the semicolon refers to covariant differentiation using the Christoffel symbol.

We shall consider  massive fermions  coupled to  topologically massive gravity (TGM) \cite{DJT} with dynamics governed by the action $S$

\be
S[ \bar \psi, \psi, e;s]= \int d^3x (\det  e) \bar \psi\left (i\gamma^\mu ( \partial_\mu + \omega_\mu + s_\mu) + m \right)\psi
+ S_{TMG}
\label{s}
\ee
where the topological massive gravity action $S_{TGM}$ reads
\be
S_{TMG} = \frac{1}{64\pi \kappa^2\eme}  S_{CS}[\omega[e]] +
\frac{1}{\kappa^2}S_{EH}[e,\omega[e]]
\label{4}
\ee
Here $\det e$ is the determinant of the dreibein fields,  and $s_\mu$  is an external source
 \be
s_\mu = -\frac14 s_{ab\mu} \gamma^a \gamma^b = -\frac{i}4  s_{ab\mu}\epsilon^{abc}\gamma^c
\ee
which transforms  covariantly under local Lorentz transformations so that $\omega + s$ transforms  as a connection.

The Chern-Simons term $S_{CS}[\omega]$ reads
\begin{equation}
S_{CS}[\omega]\,=  \int d^3x \,   \epsilon^{\mu\nu\alpha}\big(
\omega_{\mu a b} \partial_\nu\omega_{\alpha ba} + \frac{2}{3} \omega_{\mu a b}
 \omega_{\nu b c}   \omega_{\alpha c a}\big)
\label{21}
\end{equation}
and the  Einstein-Hilbert action $S_{EH}[e,\omega[e]]$ is written as
\be
S_{EH}[e,\omega[e]]
 =  \int d^3x \,   \epsilon^{\mu\nu\alpha}e_{a \mu}\big(
  \partial_\mu\omega_\alpha^a - \partial_\alpha \omega^a_\mu + \epsilon^{abc}  \omega_\mu^b\omega_\nu^c
    \big) \;.
\label{21eh}
\ee
where
\be
\omega_\alpha^a = \frac{1}{2}\epsilon^{a b c} \omega_{\alpha b c}
\ee
Note that in $d=3$ space-time dimensions the Newton constant $\kappa^{-2}$ has dimensions of mass, $[\kappa^2]=m^{-1}$, in fundamental units in which  the spin connection has dimension one, $[\omega_\mu] = 1$.  The  mass  parameter $\eme$ in $S_{CS}$ is required by dimensional consistency and can be replaced by a dimensionless effective coupling constant $\kappa^2 \eme$ which, alternatively, can be regarded as the inverse of the level $k$ of the Chern-Simons theory, $\kappa^2 \eme=1/k$. In topologically massive gravity there is a single  propagating mode with mass $\eme$ and spin $2$ \cite{DJT}.

The  generating functional $Z[s]$ of the source fields $s_\mu$  is defined by
\be
Z[s]= \int \mathcal{D}\mu_\psi \mathcal{D}\mu_G \exp(i S[ \bar \psi, \psi, e; s])
\label{z}
\ee
 where $\mathcal{D}\mu_\psi$  is the appropriate fermionic  path-integral measure defined in terms of Fujikawa variables
\cite{F1}:
\be
\mathcal{D}\mu_\psi =  \mathcal{D}\left({(\det e)^{1/2}\bar \psi}\right) \mathcal{D}\left( { (\det e)^{1/2}\psi}\right)
\label{fuji}
 \ee
Concerning the measure of the gravitational sector $\mathcal{D} \mu_G$, one could  also follow Fujikawa prescription introducing Faddeev-Popov ghosts fields  and auxiliary fields arising in the anti-ghost superfield introduced in the BRST supersymmetry approach  associated to general coordinate transformations \cite{F3}. In this respect one  should note that in ref.~\cite{Deser1} it has been shown  that TGM has no unitarity and ghost problems and is power-counting renormalizable. Moreover, arguments in \cite{enc} based on the existence  of
a functional integration measure make appear TMG  to be  finite as a quantum theory.

\subsection*{Bosonization}

As it is well-known the $d=3$ Dirac operator determinant arising from the fermionic path-integral in Eq.\eqref{z} can be written as the product of two factors according to its behavior under parity transformation \cite{AG,Kulikov}. The odd-parity contribution can be computed exactly leading to a Chern-Simons effective action
\be
 \det\left(\ i\gamma^\mu ( \partial_\mu +
\omega_\mu + s_\mu
) + m
\right)_{odd} = \exp\left( \pm\frac{i}{64\pi} S_{CS}[\omega + s]\right)
\label{det}
\ee
The parity-even contribution  can be computed within a $\partial/m$ approximation. Being the lowest parity-even term subleading to the parity-odd term, it will be disregarded in what follows.

Using this result we can write Eq.\eqref{z} as  a purely bosonic generating  functional of the form
\be
Z [s]=   \int \mathcal{D}\mu_G \exp\left(\pm   \frac{i}{64\pi} S_{CS}[\omega + s]\right)
\exp\left( \frac{i}{64\pi\kappa^2\eme} S_{CS}[\omega] + \frac{i}{\kappa^2}S_{EH}[\omega,e]\right)
\label{Z}
\ee

One can calculate from the generating functional of Eq.\eqref{Z}  the v.e.v. of the current $J^{\mu ab} $
\be
\langle J^{\mu ab} \rangle =  -\left. \frac1Z \frac{\delta Z[S]}{\delta  s_{\mu a b} } \right|_{s=0}
\ee
\be
J^{\mu\, ab} = \frac14\bar\psi \gamma^\mu [\gamma^a,\gamma^b]\psi = \frac14 e^\mu_c\bar\psi \gamma^c[\gamma^a,\gamma^b] \psi
\ee
Using eq.\eqref{Z}  the fermionic current v.e.v takes form
\bea
\langle J^{\mu ab} \rangle &=& \frac{\mp1}{64\pi} \left.\frac 1Z \int \mathcal{D}\mu_G\exp\left( \frac{i}{64\pi\kappa^2 \eme} S_{CS}[\omega] + \frac{i}{\kappa^2}S_{EH}[\omega[e],e]\right)  \left( \frac{\delta S_{CS}[\omega + s]}{\delta s_{\mu ab}}\right)
\right|_{s=0}
\\
 &=& \mp \frac{1}{256\pi}  {\epsilon^{\mu\nu\beta}} \langle ( R_{\nu\beta ab}[\tilde \omega])^{ba}\rangle_{bos}
\eea
where
\be
\langle G[\omega] \rangle_{  S_{bos}} \equiv \int \mathcal{D}\mu_G G[\omega] \exp(i S_{bos}[\omega])
\ee
Here the curvature tensor $ R_{\mu\nu ab}$  with two frame and two coordinate indices is given by
 \be R_{\mu\nu ab}[\omega]=\partial_\mu \omega_{\nu ab} - \partial_\nu\omega_{\mu ab} +\omega_{\mu ac} \omega_{\nu c b} -  \omega_{\nu ac} \omega_{\nu c b}\ee
 and the bosonized action $S_{bos}$ reads
\be
 S_{bos}  = \frac{1}{64\pi\kappa^2\eme_{\,b}} S_{CS}[\omega[e]] + \frac{1}{\kappa^2}S_{EH}[e,\omega[e]]
 \label{ref}
 \ee
 with
 \be
\eme_{\,b} = \frac\eme{1 \pm  \kappa^2 {\eme}}
\label{20}
\ee
This result implies that the Chern-Simons level of the bosonized theory is shifted from $k=1/(\kappa^2 \eme)$ to $k_b=1/(\kappa^2 \eme_b)=k\pm 1$.
Notice that if the level of the TGM is $k=1$, it is possible then to have $k_b =0$  (although formally the mass parameter $\eme_b$ diverges.) In this case the parity anomaly cancels out and  the effective action of the bosonized theory  reduces to a pure Einstein-Hilbert action in $2+1$ dimensions.

We can summarize these results establishing the following
 bosonization recipes  for  action and  current
\begin{align}
S[ \bar \psi, \psi, e;s]    \mapsto &\;  S_{bos}[e,\omega[e]]\label{recipe1}
\\
  \bar\psi \gamma^\mu [\gamma^a,\gamma^b] \psi   \mapsto &\; \mp \frac1{256\pi} {\epsilon^{\mu\nu\beta}}  R_{\nu\beta ab}[ \omega]
  \label{recipe2}
 \end{align}
We conclude that the bosonized action $S_{bos}$ corresponds to topologically massive gravity action with a modified graviton mass given by $\eme_b$ instead of the original $\eme$ so that the original graviton mass picked-up a contribution from fermion bosonization increasing or decreasing according to the choice of sign when regularizing  determinant Eq.\eqref{det}\footnote{The double sign ambiguity is inherent to any regularization in odd-dimensional spaces as can be seen by using a $\zeta$-function regularization where the sign depends on the choice of upper or lower half plane  to close the
curve where one integrates the Seeley coefficients \cite{GMSS}. The choice of a particular sign corresponds to a sign of the parity anomaly. In Pauli-Villars regularization this sign is determined by the sign of the mass of the Weyl fermion \cite{Redlich}.}.

\subsection*{The  energy-momentum ``tensor''}

One can complete  the bosonization recipe Eq.\eqref{recipe1}-Eq.\eqref{recipe2}  by relating  the  energy-momentum tensor associated to the original matter-TMG model  to the dual  action Eq.\eqref{ref}. To define the topologically massive gravity energy-momentum ``tensor''\footnote{Quotes indicate that   gauge systems with spin greater than one do not   possess gauge-invariant stress tensors, but
only integrated Poincar\'e generators \cite{DeserMac}.} we shall follow \cite{Weinberg,DJT} and consider  that the path integral measure $D\mu_G$ is restricted to   asymptotically   flat space-times. We then decompose space-time metric in the form
\be
g_{\mu\nu} = \eta_{\mu\nu} + h_{\mu\nu}
\ee
where $\eta_{\mu\nu}$ is the  flat-space-time metric with Euclidean signature and  $h_{\mu\nu}$ a deviation which is not necessarily small everywhere but vanishes at infinity.

{In the case of topologically massive gravity coupled to matter Einstein equations  can be written as
\be
G_{\alpha\beta} + \frac1\eme C_{\alpha\beta}
 = -\frac{\kappa^2}2 T_{\alpha\beta}\ee
 and  $G^{\alpha\beta}$ and $C^{\alpha\beta}$  are the  Einstein and Cotton-York tensors respectively and $T_{\alpha\beta}$ is the matter (i.e. the Dirac field) energy-momentum tensor,
\bea
G^{\alpha\beta} &=& R^{\alpha\beta} - \frac12g^{\alpha\beta} R^{\gamma}_{~\gamma}
\nonumber\\
C^{\alpha\beta} &=& \frac1{2\sqrt g} \left(\epsilon^{\alpha\rho\sigma}D_\rho R^\beta_\sigma +
\epsilon^{\beta\rho\sigma}D_\rho R^\alpha_\sigma
\right)\nonumber\\
T_{\alpha\beta} &=& \frac12 \left(\bar\psi \gamma_\beta\nabla_\alpha\psi - \overline{\nabla_\alpha \psi}\gamma_\beta \psi\right)
\eea
Here $R_{\alpha\beta}$   the Ricci curvature tensor ($R_{\alpha\beta} = g^{\gamma\delta}R_{\alpha\gamma\beta\delta}$),  $R$  the scalar curvature ($R = g^{\alpha\beta}R_{\alpha\beta} $) and
 $T_{\alpha\beta}$ the energy-momentum tensor associated with matter. We can write the previous equation as
\be
G_{\alpha\beta} +  \frac1\eme C_{\alpha\beta} + (G_{\alpha\beta}^{(1)} +  \frac1\eme C_{\alpha\beta}^{(1)}) - (G_{\alpha\beta}^{(1)} +  \frac1\eme C_{\alpha\beta}^{(1)})  = -\frac{\kappa^2}2 T_{\alpha\beta}
\nonumber \ee
where $G^{(1)}, C^{(1)}$  are the part of Einstein and Cotton-York   tensors which are linear in $h_{\alpha\beta}$.  Passing the first and third  terms in the l.h.s. to the r.h.s. one can rewrite the previous equation in the form
\be
G_{\alpha\beta}^{(1)} + \frac1\eme C_{\alpha\beta}^{(1)}= -\left( \frac{\kappa^2}2 T_{\alpha\beta} +( G_{\alpha\beta} + \frac1\eme C_{\alpha\beta}) -  (G_{\alpha\beta}^{(1)} - \frac1\eme C_{\alpha\beta} ^{(1)})
\right)
\ee
Then the quantity
\be
t_{\alpha\beta}
= \frac2{\kappa^2}
\left(
G_{\alpha\beta} -
 G^{(1)}_{\alpha\beta} + \frac1{\eme}( C_{\alpha\beta} -
 C^{(1)}_{\alpha\beta}   )
\right)
\label{chica}
\ee
can be interpreted  as the energy-momentum ``tensor'' of the gravitational field itself and then refer to
\be
\tau_{\alpha\beta} = T_{\alpha\beta} + t_{\alpha\beta}
\ee
as the total energy-momentum ``tensor'' of matter and gravitation}.

In turn, the exact equations of motion  for TGM coupled to gravity can then be written in the form
\be
 G_{\alpha\beta}^{(1)} + \frac1\eme C_{\alpha\beta}^{(1)} = -\frac{\kappa^2}{2} (T_{\alpha\beta} + t^\eme_{\alpha\beta})
\ee
with $t_{\alpha\beta}^{\!\eme}$ the energy-momentum ``tensor'' of the gravitational field for TGM with graviton mass $\eme$.
In the same vein, the equations of motion of the bosonized action Eq.\eqref{ref} can be written as
\be
 G_{\alpha\beta}^{(1)} + \frac1{\eme^b} C_{\alpha\beta}^{(1)} = -\frac{\kappa^2}{2} (t^{\eme^b}_{\alpha\beta})
\ee
Using the definition Eq.\eqref{chica}   and the fact that
the two actions are equivalent,  their equations of motion can be related through the bosonization recipe
for the matter energy-momentum tensor
\be
\frac{\kappa^2}2 T_{\alpha\beta} \to \left(\frac1{\eme_b} - \frac1\eme\right) C_{\alpha\beta}  =  \pm\frac{\kappa^2}{64\pi} C_{\alpha\beta}
\ee
which is thus mapped onto the Cotton-York tensor of the TMG.

In summary, the complete set of bosonization recipes reads
\begin{align}
S[ \bar \psi, \psi, e, \omega[e]]    \mapsto &\;  S_{bos}[e,\omega[e]]\label{recipea}
\\
  \bar\psi \gamma^\mu [\gamma^a,\gamma^b] \psi  \mapsto &\; \pm \frac1{256\pi} {\epsilon^{\mu\nu\beta}}  R_{\nu\beta ab}[ \omega]
  \label{recipeb}\\
 T_{\alpha\beta} = \frac12 \left(\bar\psi \gamma_\beta\nabla_\alpha\psi - \overline{\nabla_\alpha \psi}\gamma_\beta \psi\right)  \mapsto &\;\pm  \frac1{32\pi} C_{\alpha\beta}
 \label{recipec}
 \end{align}
Notice the important fact that these identifications do not depend explicitly on the parameters of the theory.  These results are the gravitational analog of the results of Refs. \cite{FS,LGMNS} in terms of gauge fields. As usual they should be interpreted as identities inside expectation values.

Several interesting results can be drawn form the identifications of Eqs.\eqref{ref}, \eqref{20}, \eqref{recipea}, \eqref{recipeb} and \eqref{recipec}. Let us consider the case in which $\kappa \to 0$. In this limit the TMG of Eq.\eqref{4} trivializes and the gravitational degrees of freedom reduce to diffeomorphisms of $2+1$-dimensional Minkowski  space-time. In this limit the theory reduces to a theory of free massive Dirac fermions (averaged over trivial diffeomorphisms which amount to an average over boundary conditions). From the results of Eq.\eqref{ref} and Eq.\eqref{20}, we see that, in this limit, the bosonized theory is classical. Thus, the expectation values and correlators of the stress-tensor of the Dirac field are trivial in this regime (as it is expected since if the Dirac mass is arbitrarily large there is no energy or momentum). We note that in this limit, in which gravity acts as a probe field, in order to obtain non-trivial stress energy currents it is necessary to consider a gravity theory with torsion \cite{Fr1,Le2}.
On the other hand, in the opposite  limit $\kappa \to \infty$, the TMG sector of the action drops out and the original theory is a theory of massive Weyl fermions with strongly fluctuating gravitational degrees of freedom. In this limit we see that the bosonized TMG theory now is a Chern-Simons theory for the spin connection with CS level $k_b=1$.

\vspace{.5cm}

\noindent\underline{Acknowledgments}:  We would like to thank G. Giribet for helpful comments. This work is supported in part by the National Science Foundation through the grant DMR-1064319 at
the University of Illinois, USA (EF)  and by  CONICET, ANPCYT, CIC,  and UNLP, Argentina (FAS).


\begin{thebibliography}{99}
\bibitem{Coleman}
E. Lieb and D. C. Mattis,
J.\ Math.\ Phys. {\bf 6} (1965) 304;
  S.~R.~Coleman,
  Phys.\ Rev.\ D {\bf 11} (1975) 2088;
  S. Mandelstam,
  Phys.\ Rev.\ D {\bf 11} (1975) 3026.
  \bibitem{F1}
   K.~Fujikawa,
  Phys.\ Rev.\ Lett.\  {\bf 44} (1980) 1733.

\bibitem{Roskies}
  R.~Roskies and F.~Schaposnik,
  Phys.\ Rev.\ D {\bf 23} (1981) 558.
\bibitem{Naon}
  C.~M.~Na{\'o}n,
  Phys.\ Rev.\ D {\bf 31} (1985) 2035.

  \bibitem{DJT}
  S.~Deser, R.~Jackiw and S.~Templeton,
  Ann. Phys.\  {\bf 140} (1982) 372; Erratum-ibid.\  {\bf 185} (1988) 406;
   Ann. Phys.\  {\bf 281} (2000) 409.
  \bibitem{Redlich}
  A.~N.~Redlich,
  Phys.\ Rev.\ Lett.\  {\bf 52} (1984) 18.
  \bibitem{Marino}
  E.~C.~Marino,
  Phys.\ Lett.\ B {\bf 263} (1991) 63.
   \bibitem{FS}
  E.~H.~Fradkin and F.~A.~Schaposnik,
  Phys.\ Lett.\ B {\bf 338} (1994) 253.
\bibitem{LGMNS}
 J.~C.~Le Guillou, E.~Moreno, C.~N\'u\~nez and F.~A.~Schaposnik,
  Nucl.\ Phys.\ B {\bf 484} (1997) 682
  [hep-th/9609202];
  Phys.\ Lett.\ B {\bf 409} (1997) 257
  [hep-th/9703048].
  \bibitem{AG}
  L.~Alvarez-Gaum\'e, S.~Della Pietra, G.~W.~Moore,
  Annals Phys.\  {\bf 163 } (1985)  288.
  %
  \bibitem{Ojima}
  S.~Ojima,
  Prog.\ Theor.\ Phys.\  {\bf 81 } (1989)  512.

\bibitem{Kulikov}
  I.~K.~Kulikov, P.~I.~Pronin,
  Europhys.\ Lett.\  {\bf 17 } (1992)  103-107.


\bibitem{Fr1}T.~L.~Hughes, R.~G.~Leigh and E.~Fradkin,
  Phys.\ Rev.\ Lett.\  {\bf 107} (2011) 075502
  [arXiv:1101.3541 [cond-mat.mes-hall]].
\bibitem{Le2}T.~L.~Hughes, R.~G.~Leigh and O.~Parrikar,
  Phys.\ Rev.\ D {\bf 88} (2013) 025040
  [arXiv:1211.6442 [hep-th]].
  \bibitem{V}
  M.~A.~Vasiliev,
  Phys.\ Lett.\ B {\bf 285} (1992) 225;
  In *Shifman, M.A. (ed.): The many faces of the superworld* 533-610
  [hep-th/9910096].
  Nucl.\ Phys.\ B {\bf 226} (1983) 437.
  \bibitem{A}
  O.~Aharony, G.~Gur-Ari and R.~Yacoby,
  JHEP {\bf 1212} (2012) 028
  [arXiv:1207.4593 [hep-th]].
  \bibitem{F3}
  K.~Fujikawa and O.~Yasuda,
  Nucl.\ Phys.\ B {\bf 245} (1984) 436.
  \bibitem{Deser1}
  S.~Deser and Z.~Yang,
  Class.\ Quant.\ Grav.\  {\bf 7} (1990) 1603.
    \bibitem{enc} B.~Keszthelyi and G.~Kleppe,
  Phys.\ Lett.\ B {\bf 281} (1992) 33.

  \bibitem{GMSS}
  R.~E.~Gamboa Sarav{\'\i}, M.~A.~Muschietti, F.~A.~Schaposnik and J.~E.~Solomin,
  J.\ Math.\ Phys.\  {\bf 26} (1985) 2045.


\bibitem{DeserMac}  S.~Deser and J.~G.~McCarthy,
  Class.\ Quant.\ Grav.\  {\bf 7} (1990) L119.
   \bibitem{Weinberg} S.~Weinberg, ``Gravitation and Cosmology'', J.Wiley \& Sons, New York, 1981.
\end{thebibliography}
\end{document}